\documentclass[twocolumn,superscriptaddress]{revtex4-1}

\usepackage{graphicx}
\usepackage{color}
\usepackage{soul} 
\usepackage{amsmath}
\usepackage{hyperref}
\hypersetup{colorlinks=true, citecolor=cyan, urlcolor=blue, linkcolor=blue}

\begin{document}

\title{Observation of scalable sub-Poissonian-field lasing in a microlaser}
\author{Byoung-moo Ann}
\altaffiliation[Present Address: ]{Kavli Institute of Nanoscience, Delft University of Technology, 
2628 CJ Delft, The Netherlands} 
\author{Younghoon Song}
\altaffiliation[Present Address: ]{Department of Field Application, ASML Korea, Hwaseong 18449, Korea}
\author{Junki Kim}
\altaffiliation[Present Address: ]{Department of Electrical and Computer Engineering, Duke University, Durham, North Carolina 27708, USA}
\author{Daeho Yang}
\altaffiliation[Present Address: ]{Samsung Advanced Institute of Technology, Suwon 16678, Korea}
\author{Kyungwon An}
\email{kwan@phya.snu.ac.kr}
\affiliation{Department of Physics and Astronomy \& Institute of Applied Physics, Seoul National University, Seoul 08826, Korea} 
\date{\today}

\begin{abstract}
Sub-Poisson field with much reduced fluctuations in a cavity can boost quantum precision measurements via cavity-enhanced light-matter interactions. Strong coupling between an atom and a cavity mode has been utilized to generate highly sub-Poisson fields. However, a macroscopic number of optical intracavity photons with more than 3dB variance reduction has not been possible. Here, we report sub-Poisson field lasing in a microlaser operating with hundreds of atoms with well-regulated atom-cavity coupling and interaction time. Its photon-number variance was 4dB below the standard quantum limit while the intracavity mean photon number scalable up to 600. The highly sub-Poisson photon statistics were not deteriorated by simultaneous interaction of a large number of atoms. Our finding suggests an effective pathway to widely scalable near-Fock-state lasing at the macroscopic scale.
\end{abstract}

\maketitle

\section{Introduction}
Sub-Poissonian photon sources with a reduced photon number variance\cite{Davidovich-RMP1996} are essential in quantum foundation\cite{Brune-PRL1990,Nogues}, quantum information processing\cite{Kishore-PRL2007}, quantum metrology\cite{Yurke-PRL1986,Yuen-PRL1986,Motes-PRA2016} and quantum optical spectroscopy\cite{Kalashnikov-PRX2014}. Squeezed state of light from nonlinear optical devices\cite{Loudon,Andersen}, photon-pairs from parametric down-conversion processes\cite{Smithey,Waks} or antibunched radiation from single quantum emitters\cite{Mandel-OL1979,nv-center,single-molecule,Michler-Science2000,Kuhn-PRL2002,Barros-NJP2009} are well-known examples of sub-Poissonian light sources. However, these types of light usually take place in a propagating mode and do not fit to stabilize a highly sub-Poissonian field in single cavity mode. Moreover, it has been shown that both quadrature- and amplitude-squeezing cannot exceed 3dB in a cavity by injecting externally generated squeezed light\cite{limit-of-squeezing}.

In a cavity sub-Poissonian field can play a substantial role in the study of quantum dynamics and quantum precision measurements \cite{Brune-PRL1990,Nogues,Peano-PRL2015,Korobko-PRL2017,Sayrin,Purdy,Spethmann,Braginski}. 
The cavity can enhance the matter-light coupling and allow the magnitude and phase control of the coupling so as to increase sensitivity and functionality in measurements. Moreover, it provides directional emission to enable efficient collection of signals\cite{MaKeever-Nature2003,single-atom-EP,single-molecule-sensing}. A usual approach to highly sub-Poisson cavity-field stabilization is to use coherent interaction between a single Rydberg atom and a microwave cavity\cite{Nogues,Sayrin,Weidinger-PRL1999,Rempe-PRL1990}. It can provide very strong reduction in photon number variance in the microwave region. In the optical region, however, the typical single-atom-cavity coupling is not sufficient to sustain and to stabilize an intense intracavity field due to relatively large atomic and cavity damping rates.
Toward macroscopic sub-Poissonian field stabilization, it is thus crucial to address  systems with multiple atoms in a cavity. Unfortunately, the effects of multiple atoms on the photon statistics of the cavity field have not been experimentally explored except for a few studies yielding unclear conclusions\cite{Choi-PRL2006}.

In the present work, we studied the cavity-QED microlaser\cite{An-PRL1994}, an optical analog of the micromaser\cite{Meschede-PRL1985}, operating with hundreds of atoms simultaneously in a cavity mode with near identical atom-cavity coupling and interaction time. We realized lasing of a scalable sub-Poisson field of up to 600 photons in the cavity, corresponding to an output flux of $6.2\times10^8$ photons/sec. The Mandel Q parameter\cite{Mandel-OL1979}, a normalized measure of photon number variance with respect to that of coherent light, was less than -0.6, corresponding to a photon-number variance more than 4dB below the standard quantum limit. The mean photon number and the photon statistics were well described by our extended single-atom microlaser theory. Our finding suggests that the photon number can be made further scalable while its highly sub-Poisson nature preserved or even improved by injecting more atoms at a higher speed, getting us closer to the generation of macroscopic near-Fock state fields.\cite{near-Fock-state,carmichael}.

In the quantum microlaser theory (QMT), a single-atom micromaser theory\cite{Filipowicz-PRA1986,Davidovich-RMP1996} extrapolated to many atoms, the photon number rate equation is given by $\dot{n}=G(n)-\Gamma_{\rm c} n$, where $G(n)$ is the gain function and $\Gamma_{\rm c}$ is the cavity damping rate. For both well-regulated atom-cavity interaction time $t_{\rm int}$ and coupling constant $g$, we have $G(n)=r \sin^2 (\sqrt{n+1}gt_{\rm int})$ with $r$ the injection rate of the pre-inverted two-level atoms into the cavity. The sine squared part is the probability of emitting a photon via the Rabi oscillation for an atom initially prepared in the excited state while traversing the cavity during the interaction time. Suppose now the photon number deviates from the steady-state mean photon number $\langle n \rangle (\gg 1)$ momentarily by $\delta n$, {\em i.e.,} $n=\langle n\rangle +\delta n$. Then the rate equation is reduced to $\dot{\delta n}\simeq -\left[\Gamma_{\rm c}-\frac{\partial G(n)}{\partial n} \right]_{n=\langle n\rangle} \delta n \equiv -\frac{1}{\tau} \delta n$, where $1/\tau$ is interpreted as the restoring rate of the photon number. The restoring rate for conventional lasers is less than $\Gamma_{\rm c}$ since the slope $\frac{\partial G}{\partial n}$ of the gain function, which is in the form of $G_{\rm conv}(n)=\frac{G_0 (n/n_{\rm sat})}{1+(n/n_{\rm sat})}$ \cite{Siegman-sec8.3} with $G_0$ the saturated gain and $n_{\rm sat}$ the saturation photon number, is always positive. On the other hand, for the micromaser/microlaser the restoring rate can be much larger than $\Gamma_{\rm c}$ since the gain function is oscillatory and thus it can have a negative slope. The larger restoring rate than $\Gamma_{\rm c}$ suppresses photon-number fluctuations better and thus leads to a sub-Poisson photon number distribution or a negative Mandel Q\cite{Davidovich-RMP1996}. The parameter $\tau$ appears as a correlation time in the second-order correlation function. 
Mandel Q is defined as $Q=\frac{\Delta n^2}{\langle n\rangle}-1$, where $\Delta n^2\equiv \langle n^2\rangle-\langle n\rangle^2$ is the photon number variance. For a single mode of light, Mandel Q is related to the second-order correlation at zero time delay as $g^{(2)}(0)=1+Q/\langle n\rangle$\cite{Scully-book}. We use this relation to obtain Mandel Q from the observed $g^{(2)}(0)$ and $\langle n\rangle$.

\section{Results}

\subsection{Mandel Q obtained from the second-order correlation}
\begin{figure}
\centering\includegraphics[width=0.47\textwidth]{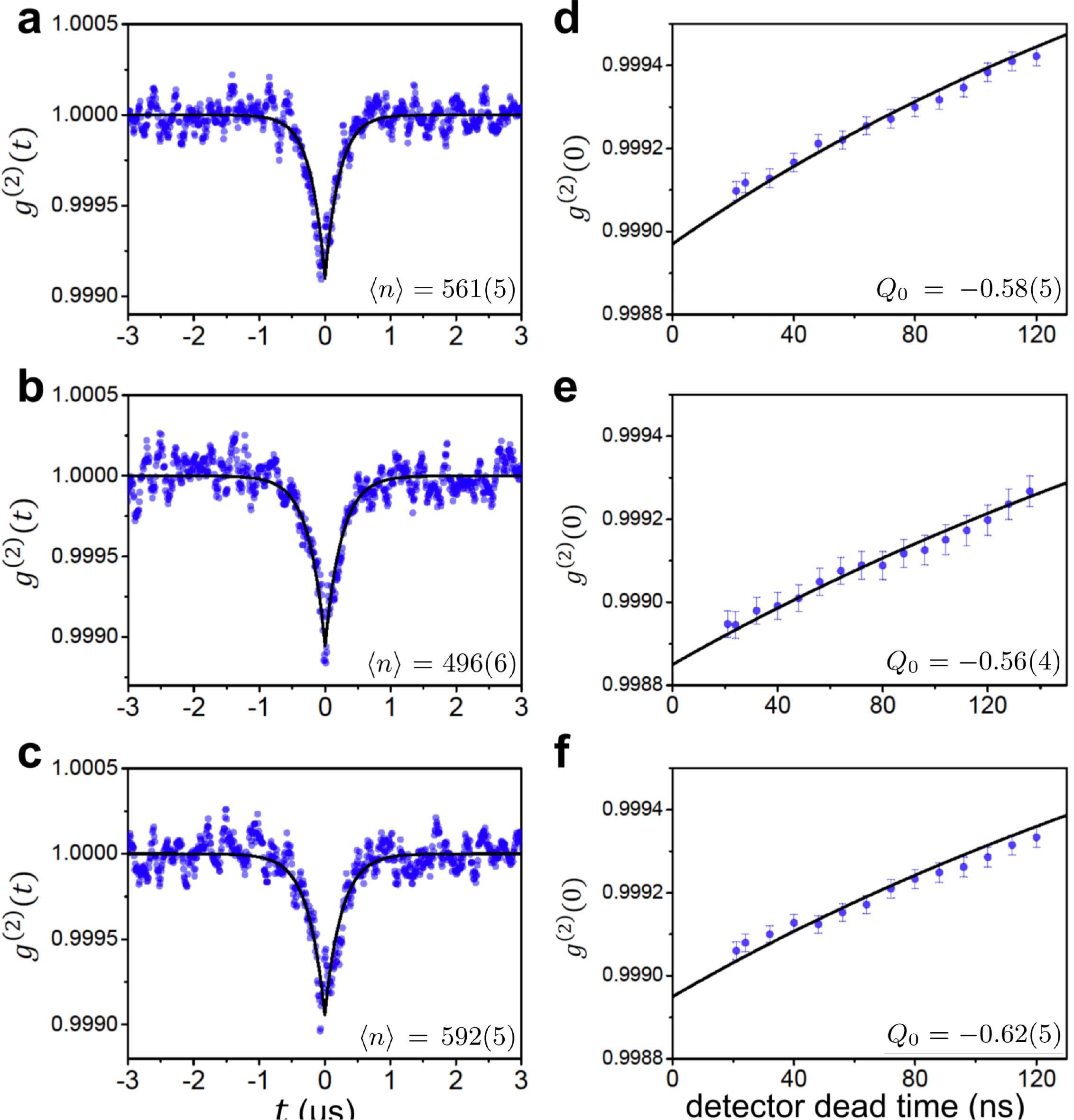}
\caption{{\bf The observed second-order correlation functions 
and the associated deadtime-free Mandel Q's.} 
(a)-(c) Observed second-order correlation function $g^{(2)}(t)$. Black curves are the fits given by $g^{(2)}(\tau)=1+\frac{Q}{\langle n\rangle} e^{-t/\tau}$. (d)-(f) Second-order correlation at zero time delay $g^{(2)}(0)$ (blue filled circles) as a function of detector deadtime. Black curves are the quadratic fits and the $y$ intercepts are deadtime-free $g^{(2)}(0)$. Experimental conditions are as follows. 
(a) $\langle N\rangle=220(10)$, $\langle n\rangle=561(5)$, $v_0=762(3)$m/s and $\Delta v/v_0=0.33$. 
(b)  $\langle N\rangle=130(9)$, $\langle n\rangle=496(6)$, $v_0=777(1)$m/s and $\Delta v/v_0=0.32$. 
(c)  $\langle N\rangle=272(14)$, $\langle n\rangle=592(5)$, $v_0=779(3)$m/s and $\Delta v/v_0=0.25$.
Here, {$\langle N\rangle$ is the intracavity mean atom number,} $v_0$ is the most probable speed of atoms and $\Delta v$ is the width (FWHM) of the velocity distribution.
Errors in $\langle N\rangle$ and $\langle n\rangle$  are the fitting error in Fig.~\ref{fig:setup}(b). Errors in $Q_0$ are mainly caused by the fitting error of $g^{(2)}(t)$ curve.
Measurement errors are indicated in parentheses (e.g. 220(10) means 220$\pm$10).
The deadtime-free Mandel Q, denoted by $Q_0$, and {the Mandel Q obtained from QMT, denoted by $Q_{\rm QMT}$,} are as follows.
(d)  $Q_0=-0.58(5)$ and $Q_{\rm QMT}=-0.719$. 
(e) $Q_0=-0.56(4)$ and $Q_{\rm QMT}=-0.698$. 
(f) $Q_0=-0.62(5)$ and $Q_{\rm QMT}=-0.781$.
}
\label{fig:g2-data}
\end{figure}

In our experiment, Mandel Q measurement was performed under five different sets of conditions. Some of the results yielding highly sub-Poisson fields with  $Q<-0.5$ are shown in Fig.~\ref{fig:g2-data}(a)-(c). Mandel Q less than -0.5 has not been reported before in the microlaser. The second-order correlation at zero time delay, $g^{(2)}(0)$, was measured with various detector deadtimes -- a finite detector deadtime deteriorates $g^{(2)}(0)$ -- as shown in Figs.~\ref{fig:g2-data}(d)-(f), using the method described by Ann {\it et al.}\cite{Ann-PRA2015}. 
By fitting the $g^{(2)}(0)$ data as a function of the detector deadtime, we then obtained the deadtime-free $g^{(2)}(0)$. Using this method, we observed deadtime-free Mandel Q (denoted by $Q_0$) less than -0.6 at a large mean photon number of 592$\pm$5 as shown in Figs.\ \ref{fig:g2-data}(c) and \ref{fig:g2-data}(f). This intracavity photon number corresponds to an output flux of $6.2\times10^8$ photons/sec, {where the output flux is given by the intracavity mean photon number in the steady state times the cavity decay rate}

The present results are clearly improved ones from those by Choi {\it et al.}\cite{Choi-PRL2006} and by Ann {\it et al.}\cite{Ann-PRA2015}, reporting Mandel Q's of -0.13 and -0.5, respectively. Here we are reporting Mandel Q less than -0.6, corresponding to reduction of photon number variance beyond the 3dB limit for the intracavity field: Mandel Q cannot go below -0.5(3dB) in a cavity by injecting externally generated squeezed light via nonlinear optical processes\cite{limit-of-squeezing}. Improving the counting electronics for the second order correlation measurement and narrowing the velocity distribution of atomic beam are main reasons for the improvement in Mandel Q results. The former is discussed by Ann {\it et al.}\cite{Ann-PRA2015} in detail. The latter is supported by the trend shown in Fig.~\ref{fig:g2-data}: we obtained the smallest Mandel Q when the velocity distribution was the narrowest. In addition to these factors, the cavity-lock electronics have been also improved so as to minimize noise signals in the second-order correlation data.

\subsection{Analysis of cavity damping during the atom-cavity interaction time} 
It should be pointed out, however, that a discrepancy around 0.15 exists between $Q_0$'s and $Q_{\rm QMT}$'s, the Mandel Q's expected from QMT. There have been several investigations regarding such discrepancies. One possible source of discrepancy is the multi-atom effect, which is known to destroy the photon-number {\em trapping} states in the micromaser\cite{Weidinger-PRL1999}. It has thus been suspected that QMT might not correctly describe the photon statistics of the micromaser as well as the microlaser working with a large number of atoms\cite{Choi-PRL2006}. However, we will show later this is not always the case.

\begin{figure}
\centering\includegraphics[width=0.47\textwidth]{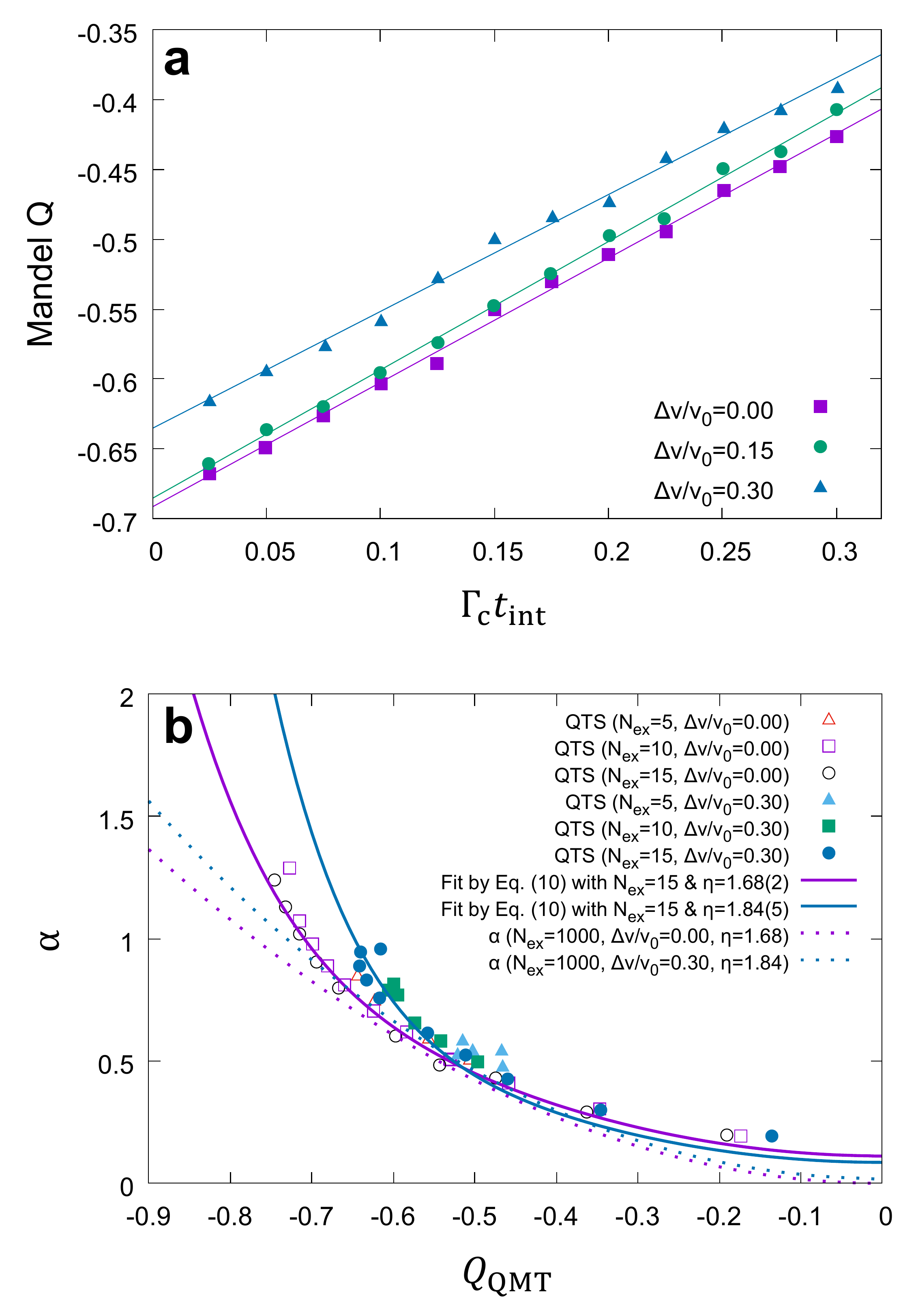} 
\caption{{\bf Accounting for cavity damping during the interaction time.} 
(a) Degradation of $Q_0$, obtained by QTS, as a function of $\Gamma_{\rm c} t_{\rm int}$ for various $\Delta v/v_0$ values. 
(b) The slope $\alpha$ in (a) as a function of $Q_{\rm QMT}$ for various $\{N_{\rm ex},\Theta,\Delta v\}$ values. For fixed $N_{\rm ex}$ and $\Delta v$, different $Q_{\rm QMT}$ values are accessed by choosing different $\Theta$. The QTS results show little dependence on $N_{\rm ex}$ in the simulation range. Solid curves indicate fits for $N_{\rm ex}=15$ cases by Eq.~(\ref{eq10}) in Methods with $\eta$ as a fitting parameter. These fits approach near-quadratic fits (dotted curves) for $N_{\rm ex} \gg 10$.}
\label{fig:QTS-results}
\end{figure}

Another possible source is the cavity damping effect. In the numerical study by Fang-Yen {\it et al.}\cite{Fang-Yen-OC2006}, quantum trajectory simulations(QTS's) including the cavity damping during the atom-cavity interaction time, which is neglected in the original QMT, resulted in Mandel Q values higher than those predicted by the QMT. This trend persisted even when the mean atom number in the cavity was less than unity, and therefore it suggested the degrade in Mandel Q was dominantly due to the damping effect rather than multi-atom effect. However, the condition of the simulation by Fang-Yen {\it et al.}\cite{Fang-Yen-OC2006} was far away from the realistic condition. Also, velocity distribution of the atomic beam was not considered in the simulation. 

For rigorous investigation of the cavity damping and multi-atom effect, we have performed extended numerical studies to cover real experiment. Our QTS results in Fig.\ \ref{fig:QTS-results}(a) show that Mandel Q linearly increases with increasing $\Gamma_{c} t_{\rm int}$ while the other system parameters $\{N_{\rm ex},\Theta, \Delta v\}$ kept fixed, where $\Theta\equiv\sqrt{N_{\rm ex}}gt_{\rm int}$, $N_{\rm ex}\equiv r\Gamma_{\rm c}^{-1}$ and $\Delta v$ the full width of the atomic velocity distribution.  These parameters fully characterize the gain function of the microlaser. We newly define $\alpha$ as the slope in Fig.\ \ref{fig:QTS-results}(a) and consider it a function of $\{N_{\rm ex}, \Theta, \Delta v\}$ in general. We then plot $\alpha$ with respect to $Q_{\rm QMT}$ as presented in Fig.\ \ref{fig:QTS-results}(b). The values of $\alpha(N_{\rm ex},\Theta, \Delta v)$ were obtained from QTS with various combinations of $\{N_{\rm ex}, \Theta, \Delta v\}$ chosen in the range $(5\le N_{\rm ex}\le 15,\; 1.5\le\Theta\le 5$ and $0\le \Delta v/v_0\le 0.30)$, which produce Mandel Q's similar to those in our experiments. Different combinations of $\{N_{\rm ex}, \Theta, \Delta v\}$ give rise to different pairs of $Q_{\rm QMT}$ and $\alpha$ but they all lie around a well defined trajectory for given $\Delta v/v_0$ in Fig.\ \ref{fig:QTS-results}(b). It suggests that $\alpha$ is approximately a function of $Q_{\rm QMT}$ only for a fixed $\Delta v/v_0$: 
\begin{equation}
Q_0 \simeq Q_{\rm QMT}+\alpha(Q_{\rm QMT})  \Gamma_{\rm c}t_{\rm int}.
\label{eq1}
\end{equation}

We investigated the semi-classical single-atom micromaser theory by Davidovich\cite{Davidovich-RMP1996}, which is the basis of QMT, and extended it to include the cavity damping effect during the atom-cavity interaction time. We could derive an explicit functional form of $\alpha(Q_{\rm QMT})$ with a dimensionless parameter $\eta$ under a weak assumption on the coarse-grain approximation (see Methods).
The solid curves in Fig.~\ref{fig:QTS-results}(b) were obtained by fitting the QTS results with $\alpha(Q_{\rm QMT})$ given by Eq.~(\ref{eq10}) in Methods with $\eta$ as a fitting parameter for the given $\Delta v/v_0$. 
Different $\Delta v/v_0$ produces different $\eta$. In the limit of large $N_{\rm ex}\gg 10$ as in the actual experiment, the $\alpha$ curves approach a parabola [dotted curves in Fig.~\ref{fig:QTS-results}(b)].

\begin{figure}
\centering\includegraphics[width=0.45\textwidth]{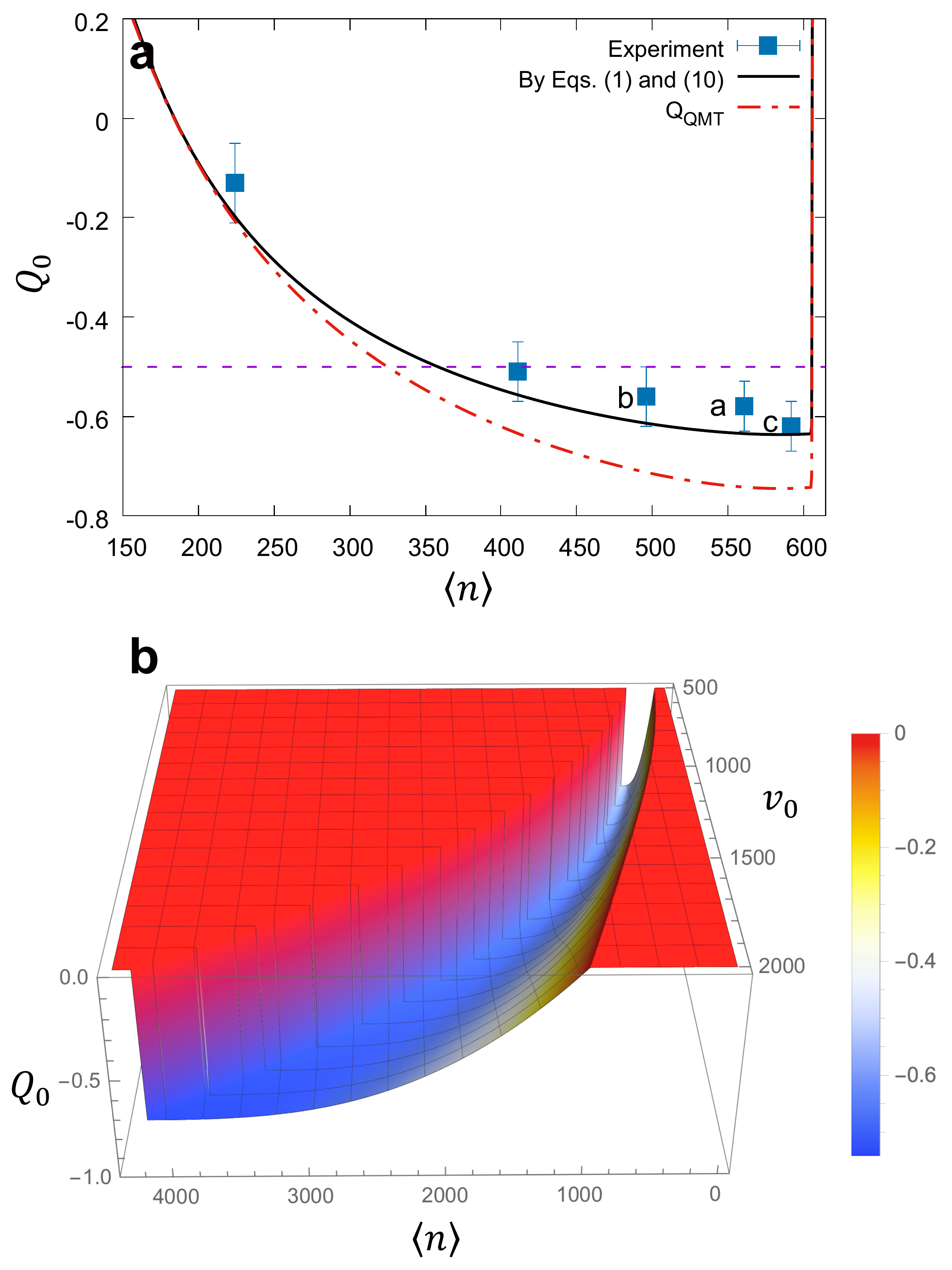} 
\caption{{\bf Scalable mean photon number with highly negative Mandel Q.} (a) Predicted $Q_0$ (black solid curve) as a function of the mean photon number $\langle n\rangle$ with a velocity-averaged gain function with $\Delta v/v_0=0.3$ and $v_0$=780m/s. The sudden increase near $\langle n\rangle\simeq 605$ is due to a quantum jump [see Fig.~\ref{fig:setup}(b)]. Equation (1) with $\alpha$'s determined in Fig.~\ref{fig:QTS-results}(b) for $N_{\rm ex}=1000$ and $\Delta v/v_0=0.3$ was used to calculate $Q_0$. For comparison, $Q_{\rm QMT}$ (red dot-dashed curve) is also shown along with the purple dashed line indicating $Q_0=-0.5$. Letters a, b and c indicate the data points from Fig.~\ref{fig:g2-data}(a)-(c), respectively.
(b) Predicted $Q_0$ as a function of both $\langle n\rangle$ and atomic velocity $v_0$ (with $\Delta v/v_0=0.3$). By scanning $v_0$ and $\langle N\rangle$ simultaneously, one can tune $\langle n\rangle$ continuously while maintaining  $Q_0<-0.6$ ($Q\rightarrow - 0.75$ as $\langle n\rangle$ approaches 4000). The cliff on the left is due to the quantum jump as in Fig.~\ref{fig:setup}(b).
}
\label{fig:scalability}
\end{figure}

In Fig.~\ref{fig:scalability}(a), we compare the experimentally observed Mandel Q ($Q_0$) with the simulation (black curve) based on Eq.~(\ref{eq1}) with the $\alpha$ (royal-blue dotted curve) determined in Fig.~\ref{fig:QTS-results}(b) for $N_{\rm ex}=1000$ and $\Delta v/v_0=0.3$, similar to the experimental values used for data in Fig.~\ref{fig:setup}. We observe good agreement between the simulation based on the extended single-atom theory and the experiment within the measurement uncertainty. The observed agreement clearly shows that the multi-atom effect is negligible on the photon statistic in our study.

\section{Discussion}

\subsection{Scalable nonclassical field beyond the 3dB limit.}
Figure \ref{fig:scalability}(a) also shows our approach is scalable in that sub-Poisson field can be generated with a mean photon number $\langle n\rangle$ scalable from 200 to 600 while maintaining negative Mandel Q. In particular, $\langle n\rangle$ is scalable over a significant range while keeping $Q_0<-0.5$. In the usual squeezing in propagating modes by nonlinear optical processes, Mandel Q cannot go below -0.5 in a cavity\cite{limit-of-squeezing}. Some of our experimental results, on the other hand, are below that limit with a large mean photon number approaching 600. The super-Poisson behavior for small $\langle n\rangle (<180)$ is due to the lasing threshold occurring near $\langle N\rangle\sim 10$ [see Fig.~\ref{fig:setup}(b)]\cite{Choi-PRL2006}. It has been shown that the lasing threshold can be eliminated by employing atoms prepared in the same superposition state\cite{single-atom-superradiance}. Using this feature the Mandel Q in the small $\langle n\rangle$ region can be further lowered.

By scanning the atomic velocity $v_0$ and the atom number $\langle N\rangle$ simultaneously, one can make the mean photon number scalable over a much wider range as illustrated in Fig.~\ref{fig:scalability}(b) while maintaining $Q_0<-0.6$ (see Fig.~\ref{fig5} in Methods for details). The largest atom number and the largest velocity are limited only by experimental capability. The intracavity atom number up to 1300 has already been demonstrated as shown in Fig.~\ref{fig:setup}(b). With a modified atomic beam source, the atom velocity can be boosted to 1500m/s\cite{new-atom-beam} and the atom number can be further increased so as to make the photon number scalable up to thousands. Using improved cavity design and atomic oven design, one can further increase the mean atom number in the cavity.

\subsection{Validity of one-atom theory.}
In Fig.~\ref{fig:scalability}(b) (also in Fig.~\ref{fig5}), the larger $\langle n\rangle$ requires the larger $\langle N\rangle$, and therefore, the validity of QMT neglecting the multi-atom effects including atom-number fluctuations might be in question. QMT fails if photon emission or absorption by any single atom affects the atom-field interaction of the other atoms significantly. Since each atom interacts with the common cavity field with a Rabi angle $\Theta_n=\sqrt{n+1}g t_{\rm int}$, the preceding statement can be rephrased as  $\Delta \Theta_n=g t_{\rm int}/2\sqrt{n+1}\ll 1$ for $\Delta n=1$ for the validity of neglecting many-atom effects\cite{An-JPSJ2003}. The lefthand side of the inequality gets even smaller as $\langle n\rangle$ and the velocity are increased (thus $t_{\rm int}$ decreased) along the valley in Fig.~\ref{fig:scalability}(b), and therefore, the multi-atom effects can be safely neglected in this approach.

\section{Methods}

\begin{figure}
\centering\includegraphics[width=0.42\textwidth]{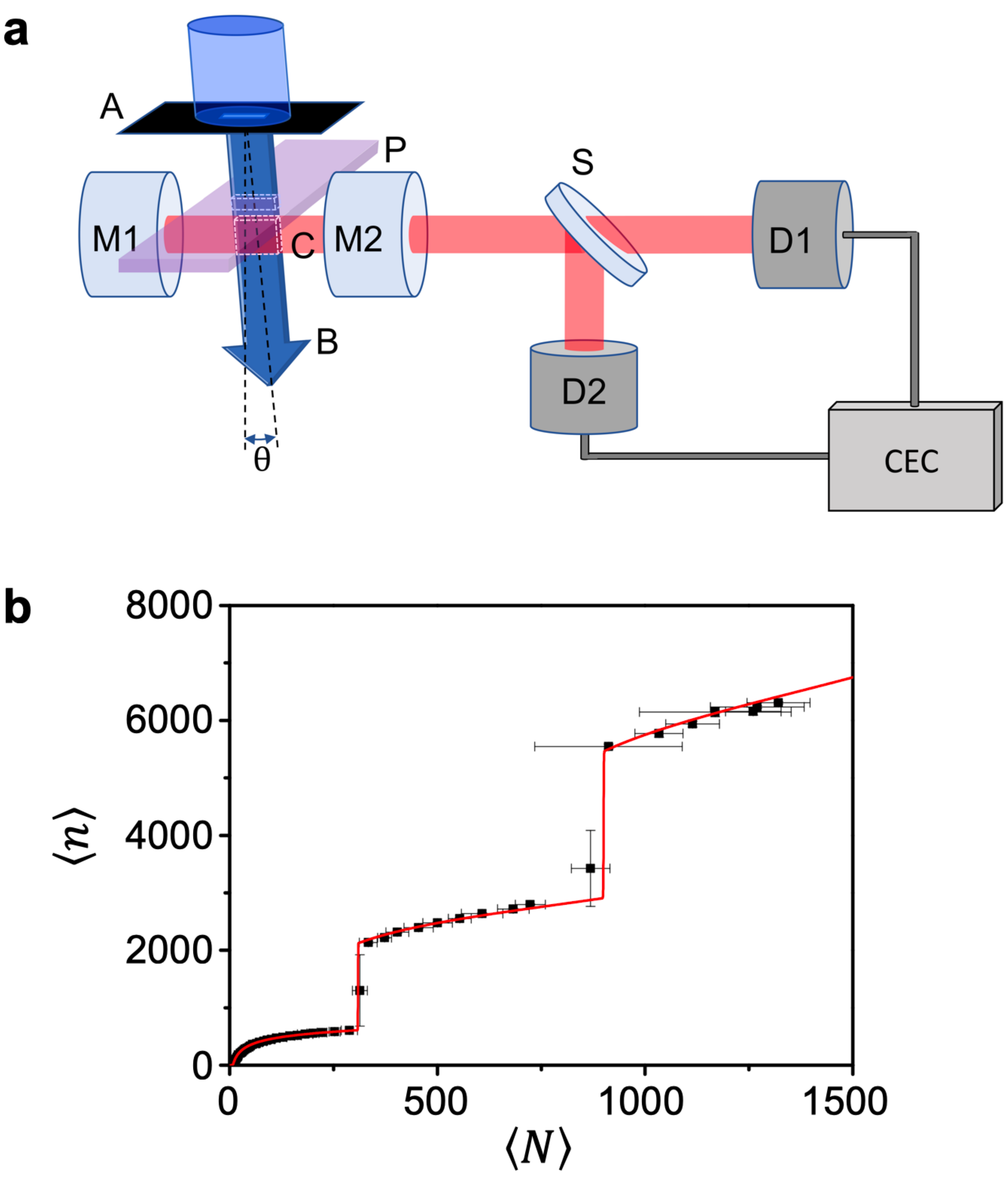}
\caption{{\bf Experimental setup and calibration method.} (a) Schematic of the cavity-QED microlaser. A: atomic beam aperture, B: atomic beam, U: unfiltered atomic beam, C: cavity mode, P: pump laser beam between A and C, M1\&M2: cavity mirrors, S: beam splitter, D1\&D2: photon-counting detector, CEC: counter electronics and computer, $\theta$: atomic beam tilt angle. The image was manually created by the authors with Microsoft Powerpoint 2016. 
(b) Observed mean photon number $\langle n \rangle$ as a function of the mean atom number $\langle N \rangle$ in the cavity. The red curve is the fit by QMT. The fit allows us to calibrate SPCM's for the microlaser output as well as the atomic fluorescence. The sudden jumps in the mean photon number occurring at $\langle N\rangle\sim 310,900$ correspond to the quantum jumps in the micromaser/microlaser\cite{Filipowicz-PRA1986,Davidovich-RMP1996,Fang-Yen-PRA2006}. 
} 
\label{fig:setup}
\end{figure}

\subsection{Experimental Setup}
Experimental schematic is shown in Fig.~\ref{fig:setup}. A Fabry-Perot type optical cavity of 1mm length forms a TEM$_{00}$ Gaussian mode, which is tuned to the resonance wavelength of $^1$S$_0\leftrightarrow ^3$P$_1$ transition of $^{138}$Ba (wavelength $\lambda = 791.1$nm, a full linewidth $\Gamma_a/2\pi=50$kHz) with a full cavity linewidth $\Gamma_{\rm c}/2\pi = 170$kHz and a mode waist $w_0$= 41$\mu$m. A supersonic barium atomic beam is collimated and made to traverse the cavity mode. The most probable speed $v_0(\simeq$780m/s) and the FWHM width $\Delta v(\simeq 0.3v_0)$ of the velocity distribution were measured from the Doppler-shifted fluorescence spectra of the atomic beam excited by a counter-propagating probe laser. Just before the atoms enter the cavity mode, they are excited by a pump laser to $^3$P$_1$ state, the upper lasing level. A collimating atomic aperture of 250$\mu$m$\times$25$\mu$m (the longer side along the cavity axis) is used to narrow the spatial distribution of the atomic beam through the cavity mode. Furthermore, the atomic beam is tilted by $\theta$=28mrad with respect to the normal incidence to the cavity mode in order to induce a traveling-wave uniform atom-cavity coupling constant\cite{An-OL1997} $\bar{g}/2\pi=190$kHz, with $\Delta g/\bar{g}=0.025$ due to the finite atomic beam size, satisfying the strong coupling condition $2\bar{g}\gg \Gamma_a, \Gamma_c$ for single atoms. The average interaction time $t_{\rm int}\equiv \sqrt{\pi}w_0/v_0\simeq 0.093\mu$s was much shorter than the atomic decay time ($1/\Gamma_a$=3.2$\mu$s) as well as the cavity decay time ($1/\Gamma_{\rm c}$=0.94$\mu$s). 

\subsection{Second-order correlation measurement setup}
The second order correlation function $g^{(2)}(\tau)$ of the microlaser output was obtained by performing Hanbury Brown-Twiss-type measurements with two single-photon count modules (SPCM's). The microlaser output was divided by a beam splitter into two and all photon arrival times in each path were recorded with a SPCM. The second-order correlation was then calculated from the photon detection records. Our scheme corresponds to a multi-start-multi-stop configuration\cite{Choi-RSI2005}. We employed a high-speed counter electronics based on field-programmable-gate-array boards to provide a synchronized clock signal to each detector and to ensure no removal of time records from counting-board-induced deadtime. The deadtime effect from intrinsic detector characteristics can be corrected by the methodology introduced by Ann {\it et al.}\cite{Ann-PRA2015}. 

\subsection{Atom and photon number calibration.}
In order to calibrate the mean atom number $\langle N\rangle$ and the mean photon number $\langle n \rangle$ in the cavity mode, we measured the fluorescence of the intracavity atoms at $^1$S$_0\leftrightarrow ^1$P$_1$ transition ($\lambda=553$nm) and the microlaser output photon flux simultaneously as the atomic beam flux was increased.  The results were then calibrated by fitting them to the distinctive theoretical curve from QMT as shown in Fig.~\ref{fig:setup}(b). This calibration method is well justified because it was proven from various studies\cite{An-JPSJ2003,Fang-Yen-PRA2006,Choi-PRL2006,Hong-PRL2012} that QMT correctly describes the mean photon number in the microlaser with a large number of atoms.

\subsection{Derivation of Eq.\ (1)}
In the semiclassical theory of the micromaser by Davidovich\cite{Davidovich-RMP1996}, the change of the photon number variance in time $T\gg t_{\rm int}$ by atomic emission is given by
\begin{eqnarray}
\frac{\delta(\Delta n^2)}{T}&=&\frac{\Delta\langle n\rangle-\Delta\langle n\rangle^2}{T} \nonumber\\
&=&r\langle P(n)\rangle + 2r\langle P(n)(n-\langle n\rangle)\rangle+r^2 \Delta P(n)^2 T, \nonumber\\
& &
\end{eqnarray}
where $\Delta P(n)^2\equiv \langle P(n )^2\rangle-\langle P(n)\rangle^2$ is the variance of $P(n)=\sin^2 (\sqrt{n+1}gt_{\rm int})$, the photon emission probability of atoms during the interaction time $t_{\rm int}$. If we assume a delta-function-like photon number distribution, the variance of $P(n)$ can be neglected and then the photon number diffusion equation in the original theory of Davidovich is recovered. In our extension, we do not neglect it since the photon number distribution has a finite width and thus $P(n)$ has a finite variance in general. In the presence of cavity decay, the right hand side would be independent of $T$ in the steady state. Based on this consideration, we replace $T$ in the last term with $t_{\rm int}$, the only time parameter in the problem with introduction of $\eta$, an unknown dimensionless factor. So, the last term becomes $2r^2 \Delta P(n)^2 \eta t_{\rm int}$. We then perform a coarse-grain approximation as
\begin{eqnarray}
\frac{\delta(\Delta n^2)}{T}&\rightarrow&\frac{d(\Delta n^2)}{dt} \nonumber\\
&=&r\langle P(n)\rangle + 2r\langle P(n)(n-\langle n\rangle)\rangle \nonumber\\
& & +2 r^2 \Delta P(n)^2 \eta t_{\rm int}.
\end{eqnarray}
Incorporating the cavity decay, we obtain
\begin{eqnarray}
\frac{d(\Delta n^2)}{dt}&=&r\langle P(n)\rangle + 2r\langle P(n)(n-\langle n\rangle)\rangle  \nonumber\\
& &-2\Gamma_{\rm c} \langle \Delta n^2\rangle+\Gamma_{\rm c}\langle n\rangle+2 r^2 \Delta P(n)^2 \eta t_{\rm int}. \nonumber\\
& &
\end{eqnarray}
The last term is our extension to Davidovich's theory. We assume a continuous and narrow photon number distribution and solve the equation for the steady state by letting $\frac{d(\Delta n^2)}{dt}=0$:
\begin{eqnarray}
0&=&rP(n_0)+2rP'(n_0)[\Delta n^2]_0+2 \eta r^2 \Delta P(n)^2 t_{\rm int}\nonumber\\
& & -2\Gamma_{\rm c}[\Delta n^2]_0+\Gamma_{\rm c} n_0,
\end{eqnarray}
where $n_0$ is the most probable photon number or the mean photon number in the cavity.
Solving for $[\Delta n^2]_0$ using $\Gamma_{\rm c} n_0 = rP(n_0)$, we get
\begin{equation}
\frac{[\Delta n^2]_0}{n_0}\left[1-\frac{r}{\Gamma_{\rm c}}P'(n_0)\right]=1+\frac{r^2 \Delta P(n_0)^2 \eta t_{\rm int}}{\Gamma_{\rm c} n_0}.
\label{eq-a5}
\end{equation}
Without the last term we have the unextended QMT result
\begin{equation}
\frac{[\Delta n^2]_{\rm QMT}}{n_0}=1+Q_{\rm QMT}=\left[1-\frac{r}{\Gamma_{\rm c}}P'(n_0)\right]^{-1}_{\rm QMT}.
\end{equation}
So, we have the following relation hold.
\begin{equation}
\left. P'(n_0)\right|_{\rm QMT}=\frac{\Gamma_{\rm c}}{r}\frac{Q_{\rm QMT}}{1+Q_{\rm QMT}}.
\label{eq8}
\end{equation}
Equation (\ref{eq-a5}) then becomes
\begin{eqnarray}
Q&=&\frac{[\Delta n^2]_0}{n_0}-1\nonumber\\
&\simeq& \left[1-\frac{r}{\Gamma_{\rm c}}P'(n_0)\right]^{-1}_{\rm QMT}\left[1+\frac{r^2 \Delta P(n_0)^2 \eta t_{\rm int}}{\Gamma_{\rm c} n_0}\right] -1\nonumber\\
&\simeq&Q_{\rm QMT}+\frac{r^2 [\Delta n]_{\rm QMT}^2  \Delta P(n_0)^2 \eta t_{\rm int}}{\Gamma_{\rm c} n_0^2} \nonumber\\
&=& Q_{\rm QMT}+\alpha \Gamma_c t_{\rm int},
\label{eq6}
\end{eqnarray}
where
\begin{equation}
\alpha \simeq \left\{\frac{r^2 [\Delta n]_{\rm QMT}^2  \Delta P(n_0)^2}{\Gamma_{\rm c}^2 n_0^2} \right\}\eta .
\label{eq10}
\end{equation}
The quantities in the curly brackets can be numerically evaluated by using the unextended QMT for the same $\Theta$ and $N_{\rm ex}$ values as those in QTS. A polynomial fit $\alpha(x)/\eta=\sum_{i=1}^{i=8} c_n x^n$ of these quantities is obtained as a function of $Q_{\rm QMT}$ and then $\eta$ is used as a fitting parameter to obtain the best fit of the QTS results of $\alpha$ in Fig.~\ref{fig:QTS-results}(b). The purple(royal blue) solid curve is the best fit obtained with $\eta=1.68\pm0.02 (\eta=1.84\pm 0.05)$ for $\Delta v/v_0=0 (\Delta v/v_0=0.3)$. These curves tend to bend upward in the region of $Q_{\rm QMT}<-0.6$. 
But this trend of bending upward diminishes as $N_{\rm ex}$ is increased toward the experimental values ($N_{\rm ex}\sim$1000) and the fit then approaches a quadratic fit [dotted curves in Fig.~\ref{fig:QTS-results}(b)] in that region by the reason discussed below.  

\begin{figure}
\centering\includegraphics[width=0.48\textwidth]{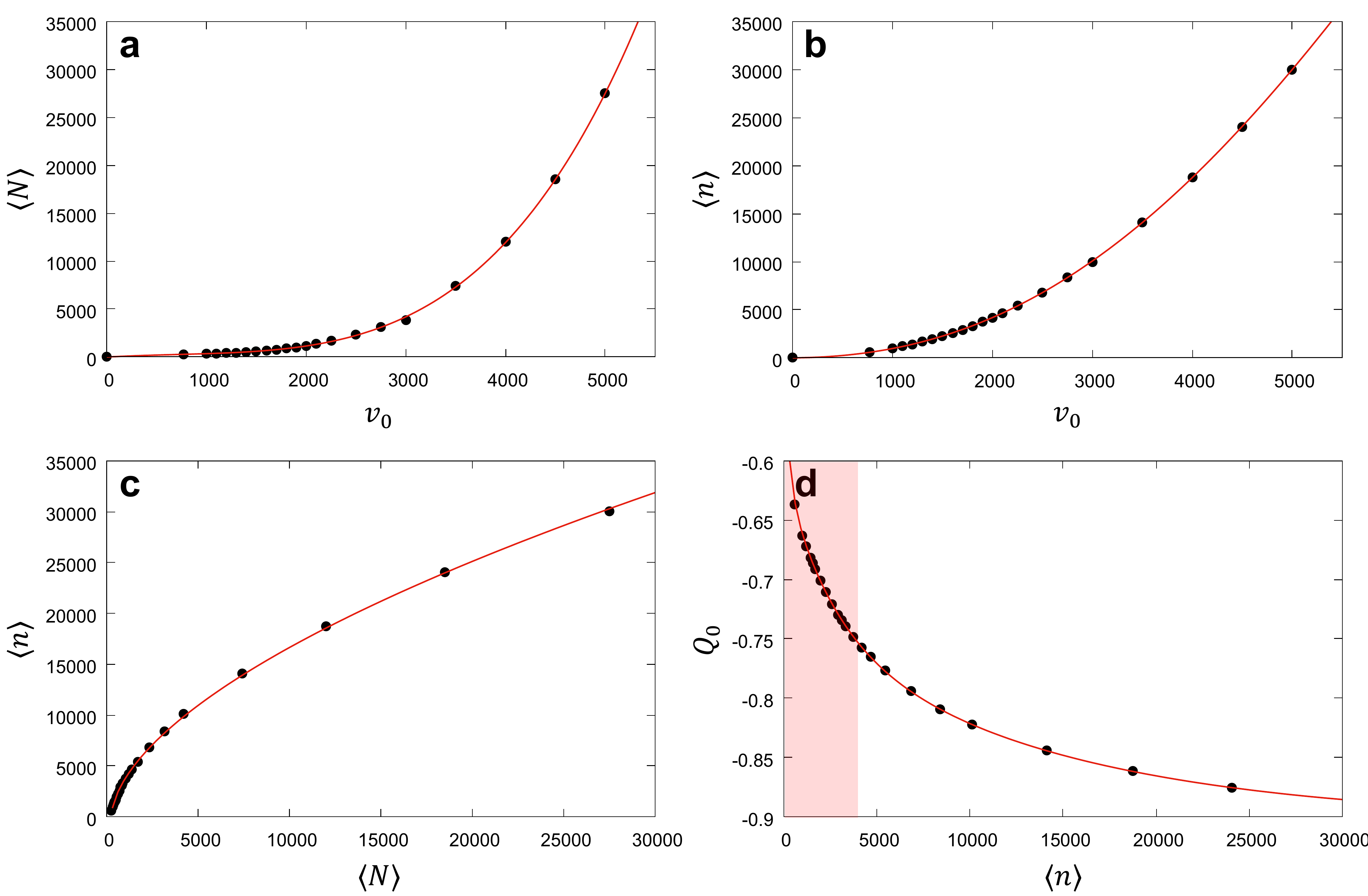}
\caption{{\bf Widely scalable mean photon number with $Q<-0.6$.}
(a) Intracavity atom number $\langle N\rangle$ corresponding the valley having minimum Mandel Q in Fig.~\ref{fig:scalability}(b) as a function of the most-probable atomic speed $v_0$.
(b) Intracavity photon number $\langle n\rangle$ corresponding to the valley as a function of $v_0$. 
(c) The resulting photon number as a function of the atom number.
(d) Predicted Mandel Q, with the correction by Eq.~(\ref{eq1}), corresponding to the valley as a function of $\langle n\rangle$.
Red solid curves are multi-exponential fits of the evaluated values (black dots). Figure \ref{fig:scalability}(b) corresponds to the shaded region in (d).
} 
\label{fig5}
\end{figure}

We can get an approximate form of $\alpha$ by expanding $\Delta P(n_0)$ in a power series of $\Delta n_0$: $\Delta P(n_0)=P'(n_0)\Delta n_0 + \frac{1}{2}P''(n_0)\Delta n_0^2 +\cdots$.
According to Eq.~(\ref{eq8}), $P'(n_0)|_{\rm QMT}$ vanishes for $Q_{\rm QMT}=0$, and thus we need to keep the higher-order terms near $Q_{\rm QMT}=0$. But for $Q_{\rm QMT}$ well away from 0, we can neglect the higher order terms and approximately have $\Delta P(n_0)\simeq P'(n_0)\Delta n_0$. To see how it comes about, consider
\begin{equation}
\frac{P''(n_0)\Delta n_0^2}{P'(n_0)\Delta n_0}\propto \frac{gt_{\rm int}}{\sqrt{n_0}}\Delta n_0\sim gt_{\rm int}. \nonumber
\end{equation}
For $\alpha$ calculation using Eq.~(\ref{eq10}), we usually fix $N_{\rm ex}$ and vary $\Theta=\sqrt{N_{\rm ex}}gt_{\rm int}$ between 2.5 and 5.  Therefore, $gt_{\rm int}=\Theta/\sqrt{N_{\rm ex}}\sim 1/\sqrt{N_{\rm ex}}\propto 1/\sqrt{n_0}$ for $N_{\rm ex}\gg1$, which is the case under our experimental condition. So
\begin{equation}
\frac{P''(n_0)\Delta n_0^2}{P'(n_0)\Delta n_0}\sim 1/\sqrt{n_0}\ll 1\;\; {\rm for}\;\; n_0\gg 1.\nonumber
\end{equation}

\hbox{}\noindent
Using this approximation, the expression for $\alpha$ can be further simplified as
\begin{eqnarray}
\alpha&\simeq& \frac{r^2 [\Delta n]_{\rm QMT}^2  [\Delta n]^2_0 P'(n_0)^2 \eta}{\Gamma_{\rm c}^2 n_0^2} \nonumber\\
&\simeq& \frac{r^2 (1+Q_{\rm QMT})  (1+Q) \left. P'(n_0)\right|_{\rm QMT}^2 \eta}{\Gamma_{\rm c}^2 } \nonumber\\
&=&\frac{(1+Q)  Q_{\rm QMT}^2  \eta}{(1+Q_{\rm QMT}) } \nonumber\\
&\simeq&\eta Q_{\rm QMT}^2 , 
\label{eq11}
\end{eqnarray}
exhibiting a quadratic dependence on $Q_{\rm QMT}$.
The dotted curves in Fig.~\ref{fig:QTS-results}(b) confirms this tendency.

\subsection{Possibility of widely scalable mean photon number with $Q$ as low as -0.9}
Highly sub-Poisson field with $Q_0$ approaching -0.9 can be obtained along the valley in Fig.~\ref{fig:scalability}(b). The velocity $v_0$ is scanned from 500m/s to 2000m/s, and for each velocity $\langle N\rangle$ is varied to obtain $\langle n\rangle$ and $Q_0$ using the QMT with the correction by Eq.~(\ref{eq1}). The resulting $Q_0$ and $\langle n\rangle$ are then plotted for various $v_0$ values. Highly sub-Poisson field with $-0.9<Q_0<-0.6$ can be obtained along the valley. The expected Mandel $Q_0$ approaches -0.9 as $\langle n\rangle \rightarrow$30,000, resulting in a macroscopic quasi Fock state. The results are shown in Fig.~\ref{fig5}.

\hbox{}

\noindent\textbf{Acknowledgements}

We thank Y.~Chough and W.~Choi for helpful discussions. 
This work was supported by Samsung Science and Technology Foundation under Project No. SSTF-BA1502- 05, the Korea Research Foundation (Grant No.~2016R1D1A109918326) and the Ministry of Science and ICT of Korea under ITRC program (Grand No.~IITP-2019-0-01402).\\





\end{document}